\documentstyle[sprocl]{article}

\def\e{\epsilon}

\def\Lpm{\Lambda^{\pm\pm}}
     
 \def\om{\omega}  \def\Om{\Omega}

\def\kp{\xi^{++}}  \def\km{\xi^{--}}  \def\kpm{\xi^{\pm\pm}}
\def\kpl{\kp_L}  \def\kml{\km_L}  \def\kpml{\kpm_L}
    
\def\kpp{\xi^{++\prime}}   \def\kmp{\xi^{--\prime}}
\def\kpmp{\xi^{\pm\pm\prime}}
\def\kppl{\kpp_L}  \def\kmpl{\kmp_L}  \def\kpmpl{\kpmp_L}

\def\tkp{{\tilde\kp}}  \def\tkm{{\tilde\km}}  \def\tkpm{{\tilde\kpm}}
\def\tkpl{\tkp_L}  \def\tkml{\tkm_L}  \def\tkpml{\tkpm_L}
    
\def\tkpp{\tilde\xi^{++\prime}}   \def\tkmp{\tilde\xi^{--\prime}}  
\def\tkppl{\tkpp_L}  \def\tkmpl{\tkmp_L}  \def\tkpmpl{\tkmp_L}

\def\oeta{\overline{\eta}}  \def\ol{\overline{\lambda}}
\def\oD{\overline{D}}

\newcommand{\p}[1]{(\ref{#1})}
\newcommand{\plabel}[1]{\label{#1}}
\newcommand{\bi}[1]{\bibitem{#1}}
\newcommand{\nn}{\nonumber\\}

\begin{document}

\title{SUPEREXTENSION $n=(2,2)$ OF THE COMPLEX LIOUVILLE EQUATION
AND ITS SOLUTION \footnote{Talk given at the XIV-th Max Born Symposium,
Karpacz, Poland, September 21-25, 1999}}

\author{A. A. KAPUSTNIKOV}

\address{Department of Physics, Dnepropetrovsk University, \\
     49050, Dnepropetrovsk, Ukraine \\
     E-mail: kpstnkv@creator.dp.ua}
\maketitle\abstracts{
It is shown that the method of the nonlinear realization of local
supersymmetry previously developed in framework of supergravity being applied
to the $n=(2,2)$ superconformal symmetry allows
one to get the new form of the exactly solvable $n=(2,2)$ super-Liouville
equation. The general
advantage of this version as compared with the conventional one is
that its bosonic part includes the complex Liouville equation. We obtain
the suitable supercovariant constraints imposed on the corresponding
superfields which provide the set of the resulting system of component
equations be the same as that in model of $N=2$, $D=4$ Green-Schwarz
superstring. The general solution of this system is derived from the
corresponding solution of the bosonic string equation.}

\section*{Introduction}

It is well-known that a doubly
supersymmetric generalization of the geometrical
approach to superstring \cite{geom,bpstv} leads in the case of $N=2$, $D=3$
Green-Schwarz superstring to the new version of the
Liouville equation referring in literature as $n=(1,1)$ \cite{bsv}. The
latter as
one can expected include the real Liouville equation in its bosonic part. The
problem, however, arise when we try to extend this result on the case of
$N=2$, $D=4$ superstring. It turns out that in this case the well-known form
of the suitable super-Liouville equation \cite{ikr}
is not relevant in virtue of absence of the complex Liouville equation in the
corresponding bosonic part.
Thus, the equation proposed in \cite{ikr} can not be applied for
description of the $N=2$, $D=4$ Green-Schwarz superstring, which as one know
is reduced to ordinary complex Liouville equation when neglecting by all
the fermionic component fields.

In this paper we would like to propose the new version of the $n=(2,2)$
super-Liouville equation which appears to be agreement with the equations of
motion of the $N=2$, $D=4$ Green-Schwarz superstring. Our approach is based
on the method of the nonlinear realization of local supersymmetries developed
by Ivanov and Kapustnikov in frame of supergravity \cite{ik}. It will be
shown that when applied to the $n=(2,2)$ superconformal symmetry this method
makes the possibility to impose the supercovariant constraints on the
superfields in such a way that all the unphysical degrees of freedom occur
in the original equation will appear
removed from the residual set of the equation of motion.
The latter amounts to the complex Liouville equation for the bosonic
worldsheet variable $\tilde{u}(\tkp,~ \tkm)$ supplemented with two first order
free equations $\tilde{\partial}_{--}\lambda^+(\tkp,~ \tkm) =
\tilde{\partial}_{++}\lambda^-(\tkp,~ \tkm) = 0$ for the fermions of opposite
chirality.

In Section 3 we present the general solution of this equation in terms
of the restricted Lorentz harmonic variables \cite{ghs}, which by a
proper fashion extends a
corresponding bosonic string solution obtained in \cite{biku}.

\section{New version of the $n=(2,2)$ super-Liouville equation}

\subsection{Linear realization}

We begin with the linear
realization of two copies of one dimensional superconformal group acting
separately on the light-cone complex coordinates of $N=2$, $D=4$ superstring
${\bf C}^{(2 \mid 2)} =
(\kpl = \kp + i\eta^+\oeta^+, \eta^+; \kml =
\km + i\eta^-\oeta^-, \eta^-)$:
\begin{eqnarray}\plabel{csct}
\kpmpl &=& \Lpm - \oeta^{\pm}\oD_{\pm}\Lpm  \\
&=& a_L^{\pm\pm}(\kpml) +
2i\eta^{\pm}\overline{\e}^{\pm}(\kpml)g^{(\pm\pm)}(\kpml)e^{i\rho^{(\pm\pm)}(\kpml)}, \nn
\eta^{\pm\prime} &=& -\frac{i}{2}\oD_{\pm}\Lpm
= \e^{\pm}(\kpml) + \eta^{\pm}g^{(\pm\pm)}(\kpml)e^{i\rho^{(\pm\pm)}(\kpml)}, \nn
a_L^{\pm\pm}(\kpml) &=& \kpml + a^{\pm\pm}(\kpml) +
i\e^{\pm}(\kpml)\overline{\e}^{\pm}(\kpml), \nn
g^{(\pm\pm)} &=& \sqrt{1 + \partial_{\pm\pm}a^{\pm\pm} +
i(\e^{\pm}\partial_{\pm\pm}\overline{\e}^{\pm}  +
\overline{\e}^{\pm}\partial_{\pm\pm}\e^{\pm})}.   \nonumber
\end{eqnarray}
In Eq. \p{csct} the general superfield (SF)
\begin{equation}\plabel{Lp}
\Lpm(\kpml, \eta^{\pm}, \oeta^{\pm}) = a_L^{\pm\pm}(\kpml) +
2i\eta^{\pm}\overline{\e}^{\pm}(\kpl)g^{(\pm\pm)}(\kpml)e^{i\rho^{(\pm\pm)}(\kpl)}
\end{equation}
$$
+ 2i\oeta^{\pm}\e^{\pm}(\kpml) -
2i\eta^{\pm}\oeta^{\pm}g^{(\pm\pm)}(\kpml)e^{i\rho^{(\pm\pm)}(\kpml)},
$$
is composed out from parameters $\e^+(\kpl)$, $\e^-(\kml)$ of local
supertranslations; two real parameters $a^{++}(\kpl)$, $a^{--}(\kml)$ of
$D1$-reparametrizations and two real parameters
$\rho^{++}(\kpl)$, $\rho^{--}(\kml)$ describing
local $U(1) \times U(1)$-rotations.
The
spinor covariant derivatives are defined as
\begin{eqnarray}\plabel{ccovD}
D_{\pm} &=& \partial_{\pm} +
2i\oeta^{\pm}\partial_{\pm\pm},   \\
\oD_{\pm} &=& \overline{\partial}_{\pm}.   \nonumber
\end{eqnarray}
It is worth to mention that since the parameters $\kpmpl$ and
$\eta^{\pm\prime}$
in Eqs. \p{csct} are
subjected to the constraints
\begin{equation}\plabel{cDel}
D_{\pm}\xi^{\pm\pm\prime}_L -
2i\oeta^{\pm\prime}D_{\pm}\eta^{\pm\prime} = 0
\end{equation}
the {\it flat} spinor covariant derivatives \p{ccovD}
are transformed homogeneously with respect to \p{csct}
\begin{equation}\plabel{tpccovDL}
D_{\pm} = (D_{\pm}\eta^{\pm\prime})D_{\pm}^{\prime}.
\end{equation}
Therefore, the following superconformal-covariant equation can be
proposed as a natural candidate for $n=(2,2)$ superextension of
the corresponding $n=(1,1)$ super-Liouville equation \cite{bsv}
\begin{equation}\plabel{nvn22sl}
D_-D_+W = e^{2W}\Psi^{--}_+\Psi^{++}_-.
\end{equation}
In Eq. \p{nvn22sl} one double-analytical SF
\begin{equation}\plabel{W}
W(\kpml, \eta^{\pm})
= u(\kpml) + \eta^+\psi^-(\kpml) + \eta^-\psi^+(\kpml) +
\eta^-\eta^+F(\kpml),
\end{equation}
and two general SFs $\Psi_+(\kpl, \eta^+, \oeta^+)$,
$\Psi_-(\kml, \eta^-, \oeta^-)$, depending separately on the
$(2,0)$ and $(0,2)$ light-cone variables, are introduced.
\footnote{We omit temporarily the upper indices of SFs $\Psi$ and $M$ for the
enlightening of formulas but we shall come back to them in Section 3.}
Eq. \p{nvn22sl} is invariant under the following gauge transformations
\begin{eqnarray}\plabel{WPsitrpr}
W^{\prime}(\kpmpl, \eta^{\pm\prime}) &=& W(\kpml, \eta^{\pm}) -
\frac{1}{2}ln(\oD_+\oeta^{+\prime}) -
\frac{1}{2}ln(\oD_-\oeta^{-\prime}), \\
\Psi^{\prime}_+(\kppl, \eta^{+\prime}, \oeta^{+\prime}) &=&
(D_+\eta^{+\prime})^{-1}(\oD_+\oeta^{+\prime})
\Psi_+(\kpl, \eta^+, \oeta^+), \nn
\Psi^{\prime}_-(\kmpl, \eta^{-\prime}, \oeta^{-\prime}) &=&
(D_-\eta^{-\prime})^{-1}(\oD_-\oeta^{-\prime})
\Psi_-(\kml, \eta^-, \oeta^-).
\nonumber
\end{eqnarray}
Note that
due to the nilpotence of the covariant derivatives $(D^2_{\pm}=0)$
the SFs $W$ and $\Psi_{\pm}$ included in the Eq. \p {nvn22sl} appear
restricted
\begin{equation}\plabel{constr}
D_{\pm}\Psi_{\pm} + 2(D_{\pm}W)\Psi_{\pm} = 0.
\end{equation}
A particular property we shall encounter with here is, however, that the
constraints \p{constr} can be solved explicitly in terms of the
{\it unrestricted} Fs
\begin{equation}\plabel{sol}
\Psi_+ = D_+M + 2(D_+W)M, \qquad \Psi_- = D_-N + 2(D_-W)N.
\end{equation}
In Eq. \p{sol} the general SFs
\footnote{It can be shown that in the
case of chiral SFs $M(\kpl, \eta^+), N(\kml, \eta^-)$ Eq. \p{nvn22sl}
is reduced
to free one $D_+D_-\tilde{W} = 0$ for the SF $\tilde{W} = W + \frac{1}{2}ln(MN)$.}
\begin{eqnarray}\plabel{MN}
M(\kpl, \eta^+, \oeta^+) &=&
f(\kpl) + \eta^+\om^-(\kpl) +  \\
&& \oeta^+\overline{\chi}^-(\kpl) +
\eta^+\oeta^+m^{--}(\kpl),  \nn
N(\kml, \eta^-, \oeta^-) &=&
g(\kml) + \eta^-\om^+(\kml) + \nn
&& \oeta^-\overline{\chi}^+(\kml) +
\eta^-\oeta^-n^{++}(\kml),\nonumber
\end{eqnarray}
are supposed transform as a superconformal densities
\begin{eqnarray}\plabel{MNtr}
M^{\prime}(\kppl, \eta^{+\prime}, \oeta^{+\prime}) &=&
(\oD_+\oeta^{+\prime})M(\kpl, \eta^+, \oeta^+), \\
N^{\prime}(\kmpl, \eta^{-\prime}, \oeta^{-\prime}) &=&
(\oD_-\oeta^{-\prime})N(\kml, \eta^-, \oeta^-). \nonumber
\end{eqnarray}
Although the component content of SFs $W,M,N$ even upon the gauge fixing is
still too large to be related with the $N=2$, $D=4$ superstring there is very
important feature of Eq. \p{nvn22sl}. It contains
{\it complex} Liouville equation in its bosonic part
\begin{equation}\plabel{LEq}
\partial_{++}\partial_{--}u(\kpml) =
\frac{1}{4}e^{2u(\kpml)}m^{--}(\kpl)n^{++}(\kml) + ...,
\end{equation}
where all the unessential terms in the r.h.s. are omitted. It is clear,
however, that to be connected with the superstring theory
the SFs we have considered here must be covariantly
constrained. In the next Section we are going to show that the desirable
constraints could be imposed in frame of the
nonlinear realization of $n=(2,2)$ superconformal symmetry in which the
original SFs becomes reducible.

\subsection{Nonlinear realization}
To see this let us suppose that the v.e.v. of the component fields
$m^{--}(\kpl)$ and $n^{++}(\kml)$ in \p{MN} are not
equal to zero and as consequence of this the local supersymmetry \p{csct} is
actually spontaneously broken. In this case the fermionic components
$\chi^{\pm}$ acquire the sense
of the corresponding Goldstone fermions and one can exploit them
for the singling out of the complex Liouville equation from the system
\p{nvn22sl} in a manifestly covariant manner. Indeed, it is well-known that
in the models with spontaneously broken supersymmetry all the SFs
becomes reducible \cite{eiak}, \cite{ik}. Their irreducible parts are
transformed, however, universally with respect to the action of the original
supergroups, as the linear representations of the underlying
unbroken subgroups but with the parameters depending nonlinearly on the
Goldstone fermions. It makes the possibility to impose generally on the SFs
in question some absolutely covariant restrictions providing to remove
out from the model under consideration undesirable degrees of
freedom. Here we can to avail oneself of the opportunity to restrict the
SFs enter the Eq. \p{nvn22sl} with the help of this approach.

For the beginning let us derive the
nonlinear realization of the superconformal symmetry in superspace. Following
closely to the general method developed in \cite{cwz}, \cite{ik}
we need firstly splits the general finite element of the group \p{csct}
\begin{equation}\plabel{G}
G(\zeta_L) \equiv \zeta_L^{\prime},
\end{equation}
where
$\zeta_L=\{\kpml, \eta^{\pm}\}$, onto the
product of two successive transformations
\begin{equation}\plabel{suctr}
G(\zeta_L) = K(G_0(\zeta_L)).
\end{equation}
In Eq. \p{suctr} the following standard notations are used. As before the
$G_0(\zeta_L)$ refer to the "primes" coordinates $\zeta_L^{\prime}$ but index
{\it zero} means that they referring now only to the stability subgroup
\begin{eqnarray}\plabel{Gvac}
\kpmpl &=& \kpml + a^{\pm\pm}(\kpml), \\
\eta^{\pm\prime} &=& \eta^{\pm}e^{i\rho^{(\pm\pm)}(\kpml)}\sqrt{1 +
\partial_{\pm\pm}a^{\pm\pm}}.
\nonumber
\end{eqnarray}
The latter include only the ordinary conformal transformations (parameters
$a^{\pm\pm}(\kpml)$) supplemented with the local $U(1) \times U(1)$-rotations
(parameters $\rho^{(\pm\pm)}(\kpml)$). Note, that
the first multiplier in the decomposition \p{suctr} is easily recognized as
the representatives of the left coset space $G/G_0$
\footnote{In virtue of \p{suctr} all the parameters in
\p{csct} should be regarded as composite ones which are composed out from the
parameters of transformations \p{Gvac} and \p{coset}.}
\begin{eqnarray}\plabel{coset}
K^{\pm\pm}(\zeta_L) &=&
\kpml + i\e^{\pm}(\kpml)\overline{\e}^{\pm}(\kpml) \nn
&& + 2i\eta^{\pm}\overline{\e}^{\pm}(\kpml)
\sqrt{1 + i(\e^{\pm}\partial_{\pm\pm}\overline{\e}^{\pm}  +
\overline{\e}^{\pm}\partial_{\pm\pm}\e^{\pm})},   \\
K^{\pm}(\zeta_L) &=& \e^{\pm}(\kpml) + \eta^{\pm}
\sqrt{1 + i(\e^{\pm}\partial_{\pm\pm}\overline{\e}^{\pm}  +
\overline{\e}^{\pm}\partial_{\pm\pm}\e^{\pm})}.
\nonumber
\end{eqnarray}
It deserves to mention that in the decomposition \p{suctr} the
comultipliers $K$ and $G_0$ are chosen in such a way that the irreduciblity
constraint \p{ccovD} is satisfied separately for both of them. The
prescription for constructing the corresponding nonlinear realization is as
follows \cite{ik}. Let us identify the local parameters
$\e^{\pm}(\kpml)$, $\overline{\e}^{\pm}(\kpml)$ in \p{coset} with the Goldstone fields
$\lambda^{\pm}(\kpml)$, $\ol^{\pm}(\kpml)$
\begin{eqnarray}\plabel{cosetNR}
\tilde{K}^{\pm\pm}(\tilde{\zeta}_L) &=&
\tkpml + i\lambda^{\pm}(\tkpml)\ol^{\pm}(\tkpml) \nn
&&+ 2i\tilde{\eta}^{\pm}\ol^{\pm}(\tkpml)
\sqrt{1 + i(\lambda^{\pm}\tilde{\partial}_{\pm\pm}\ol^{\pm}  +
\ol^{\pm}\tilde{\partial}_{\pm\pm}\lambda^{\pm})},   \\
\tilde{K}^{\pm}(\tilde{\zeta}_L) &=& \lambda^{\pm}(\tkpml) + \tilde{\eta}^{\pm}
\sqrt{1 + i(\lambda^{\pm}\tilde{\partial}_{\pm\pm}\ol^{\pm}  +
\ol^{\pm}\tilde{\partial}_{\pm\pm}\lambda^{\pm})}
\nonumber
\end{eqnarray}
and take for $\tilde{K}(\tilde{\zeta}_L)$ the transformation law associated to
\p{suctr}
\begin{equation}\plabel{NR}
G(\tilde{K}(\tilde{\zeta}_L)) = \tilde{K}^{\prime}(\tilde{G}_0(\tilde{\zeta}_L)).
\end{equation}
In Eq. \p{NR} the newly introduced coordinates
$\tilde{\zeta}_L=\{\tkpml, \tilde{\eta}^{\pm}\}$ are transformed differently as
compared with $\zeta_L=\{\kpml, \eta^{\pm}\}$ in \p{csct}. Indeed, in accordance
with \p{Gvac} they change only under the vacuum stability subgroup
\begin{eqnarray}\plabel{GvacNR}
\tkpmpl &=& \tkpml + \tilde{a}^{\pm\pm}(\tkpml), \\
\tilde{\eta}^{\pm\prime} &=& \tilde{\eta}^{\pm}
e^{i\tilde{\rho}^{(\pm\pm)}(\tkpml)}\sqrt{1 + \tilde{\partial}_{\pm\pm}\tilde{a}^{\pm\pm}},
\nonumber
\end{eqnarray}
where the parameters $\tilde{a}^{\pm\pm}(\tkpml)$ and $\tilde{\rho}^{(\pm\pm)}(\tkpml)$
turns out to be dependent nonlinearly on the fields
$\lambda^{\pm}(\kpml)$, $\ol^{\pm}(\kpml)$. Eqs. \p{NR} and \p{GvacNR}
determine the
transformation properties of the Goldstone fermions
$\lambda^{\pm}(\kpml)$, $\ol^{\pm}(\kpml)$ with respect to the
nonlinear realization of the superconformal group $G$ in coset space
\p{cosetNR}.

\section{Splitting superspace and irreducible form of SFs}

Up to now we have dealt with only formal prescription of construction of the
nonlinear realization of superconformal group $G$ without any relation
of this procedure to the original equation \p{nvn22sl}. Nevertheless, there is
the simple possibility to gain a more deeper insight into the model we started
with if we compare two Eqs. \p{G} and \p{NR}. We find that
$\tilde{K}(\tilde{\zeta}_L)$ transform under $G$ in precisely the same manner as the
initial coordinates $\zeta_L$ of superspace ${\bf C}^{(2 \mid 2)}$. Thus we
have the unique possibility to identify them
\begin{equation}\plabel{CV}
\zeta_L = \tilde{K}(\tilde{\zeta}_L).
\end{equation}
Eq. \p{CV} establish the relationship between two forms of the realization of
superconformal symmetries in superspace. One of the remarkable futures of the
transformations \p{CV} is that superspace of the nonlinear realization
${\bf \tilde{C}}^{(2 \mid 2)}=\tilde{\zeta}_L$ turns out to be completely "splitting"
in virtue of the transformations \p{GvacNR} which are not mixed the bosonic
and fermionic variables. Due to this very important fact the SFs of the
nonlinear realization becomes actually reducible. Indeed, let us perform the
change of variables \p{CV} in the Eq. \p{nvn22sl}
\begin{equation}\plabel{tnvn22sl}
\tilde{D}_-\tilde{D}_+\tilde{W} =
e^{2\tilde{W}}\tilde{\Psi}_+\tilde{\Psi}_-,
\end{equation}
where the SFs and covariant
derivatives of the nonlinear realization \p{NR}, \p{GvacNR} and \p{CV}
are introduced
\begin{equation}\plabel{tW}
W = \tilde{W} - \frac{1}{2}ln(\overline{\tilde{D}}_+\oeta^+) -
\frac{1}{2}ln(\overline{\tilde{D}}_-\oeta^-), \qquad
D_{\pm} = (\tilde{D}_{\pm}\eta^{\pm})^{-1}\tilde{D}_{\pm},
\end{equation}
\begin{equation}\plabel{tsol}
\tilde{\Psi}_+ = \tilde{D}_+\tilde{M} + 2(\tilde{D}_+\tilde{W})\tilde{M}, \qquad
\tilde{\Psi}_- = \tilde{D}_-\tilde{N} + 2(\tilde{D}_-\tilde{W})\tilde{N},
\end{equation}
$$
M(\kpl, \eta^+, \oeta^+) =
(\overline{\tilde{D}}_+\oeta^+)
\tilde{M}(\tkpl, \tilde{\eta}^+, \overline{\tilde{\eta}}^+),
$$
\begin{equation}\plabel{tMN}
N(\kml, \eta^-, \oeta^-) =
(\overline{\tilde{D}}_-\oeta^-)
\tilde{N}(\tkml, \tilde{\eta}^-, \overline{\tilde{\eta}}^-).
\end{equation}
Note should be taken that the covariant derivatives $\tilde{D}_{\pm}$ in
\p{tnvn22sl} have the same structure as those of linear realization
\p{ccovD}. This follows from the structure of the coset space
representatives \p{coset} which are defined in such a way that the
irreducibility conditions \p{cDel} are fulfilled for them automatically.

Although the form of the Eq. \p{tnvn22sl} is precisely the same as the
original one \p{nvn22sl} the SFs of the nonlinear realization appearing in
\p{tnvn22sl} are distinguished drastically from the SFs of linear realization.
As it follows from \p{GvacNR} and \p{tpccovDL} the SFs $\tilde{W}$ and $\tilde{\Psi}$
are transformed under the action of $G$ only with respect to their stability
subgroup \p{GvacNR}
\begin{eqnarray}\label{tWPsitrpr}
\tilde{W}^{\prime}(\tkpmpl, \tilde{\eta}^{\pm\prime}) &=&
\tilde{W}(\tkpml, \tilde{\eta}^{\pm}) -
\frac{1}{2}ln(\overline{\tilde{D}}_+\overline{\tilde{\eta}}^{+\prime}) -
\frac{1}{2}ln(\overline{\tilde{D}}_-\overline{\tilde{\eta}}^{-\prime}), \\
\tilde{M}^{\prime}(\tkppl, \tilde{\eta}^{+\prime}, \overline{\tilde{\eta}}^{+\prime}) &=&
(\overline{\tilde{D}}_+\overline{\tilde{\eta}}^{+\prime})
\tilde{M}(\tkpl, \tilde{\eta}^+, \overline{\tilde{\eta}}^+), \nn
\tilde{N}^{\prime}(\tkmpl, \tilde{\eta}^{-\prime}, \overline{\tilde{\eta}}^{-\prime}) &=&
(\overline{\tilde{D}}_-\overline{\tilde{\eta}}^{-\prime})
\tilde{N}(\tkml, \tilde{\eta}^-, \overline{\tilde{\eta}}^-).
\nonumber
\end{eqnarray}
Substituting here the explicit form of gauge parameters deduced from the
transformations \p{GvacNR}
\begin{equation}\plabel{Deta}
\overline{\tilde{D}}_{\pm}\overline{\tilde{\eta}}^{\pm\prime} =
e^{i\tilde{\rho}^{(\pm\pm)}(\tkpml)}\sqrt{1 +
\tilde{\partial}_{\pm\pm}\tilde{a}^{\pm\pm}(\tkpml)},
\end{equation}
one concludes that all the component fields of the SFs $\tilde{W}$ and $\tilde{M},
\tilde{N}$ are transformed {\it independently} of each other. Thus we can put down
the following manifestly covariant constraints
\begin{equation}\plabel{Wconstr}
\tilde{W}(\tkpml, \tilde{\eta}^{\pm})
= \tilde{u}(\tkpml),
\end{equation}
$$
\tilde{M}(\tkpl, \tilde{\eta}^+, \overline{\tilde{\eta}}^+) =
\tilde{\eta}^+\overline{\tilde{\eta}}^+\tilde{m}^{--}(\tkpl),
$$
\begin{equation}\plabel{MNconstr}
\tilde{N}(\tkml, \tilde{\eta}^-, \overline{\tilde{\eta}}^-) =
\tilde{\eta}^-\overline{\tilde{\eta}}^-\tilde{n}^{++}(\tkml).
\end{equation}
which leaves intact the $G$-invariance of theory. Returning these constraints
back into the Eq. \p{tnvn22sl} we obtain the final component form of the Eq.
\p{nvn22sl}
\begin{equation}\plabel{nLEq}
\tilde{\partial}_{--}\tilde{\partial}_{++}\tilde{u} =
\frac{1}{4}e^{2\tilde{u}}\tilde{m}^{--}(\tkpl)\tilde{n}^{++}(\tkml).
\end{equation}
This Eq. together with chirality conditions of the Goldstone fermions
$\lambda^{\pm}(\kpml)$, $\ol^{\pm}(\kpml)$
gives the whole system of Eqs. describing dynamics of the $N=2$, $D=4$
superstring in the component level \cite{fmb}.

\section{General solution}
Let us consider shortly the problem of construction of general solution
of the Eq. \p{nvn22sl}. It is well-known that the Virasoro constraints
simplifying significantly
the string equations of motion can generally be solved in terms of two copies
(left- and right-moving) of the Lorentz harmonic variables
parameterizing  the compact coset spaces isomorphic to the
$(D-2)$-dimensional sphere \cite{fmb}
\begin{equation}\plabel{cs}
  S_{D-2}={SO(1,D-1)
\over SO(1,1) \times SO(D-2) \times K_{D-2}}
\end{equation}
Moreover, it was shown in \cite{biku} that from these variables
the particular Lorentz covariant combinations can be formed which
resolve generally the corresponding nonlinear $\sigma$-model equations of
motion inspired by the bosonic strings in the geometrical approach
\cite{geom}.
By the construction the number of two copies of chiral variables
parameterizing the coset space \p{cs} is apparently enough to recover the
$2(D-2)$ physical degrees of freedom of $D$-dimensional bosonic strings.
But in the case of superstrings these variables replaced by the
world-sheet superfields must be properly restricted to provide the necessary
balance between bosonic
and fermionic degrees of freedom $(D-2)B~\equiv~(D-2)F$.

In this Section we shall show that the suitable constraints can be achieved
within the method of the  nonlinear realization of superconformal symmetry
developed in Section 1.2.

Proceeding from \cite{biku} one can check that the general solution of the
Liouville Eq. \p{LEq} can be written in form
\begin{eqnarray}\plabel{LEqsol}
e^{-2\tilde{u}(\tkpml)} &=& \frac{1}{2}\tilde{r}_m^{++}(\tkml)\tilde{l}^{--m}(\tkpl), \\
\tilde{m}^{--}_{++}(\tkpl) &=&
\tilde{l}_m^{--}(\tkpl)\tilde{\partial}_{++}\tilde{l}^m(\tkpl), \nn
\tilde{n}^{++}_{--}(\tkml) &=&
\tilde{r}_m^{++}(\tkml)\tilde{\partial}_{--}\tilde{r}^m(\tkml),
\nonumber
\end{eqnarray}
where the left(right)-moving Lorentz harmonics are normalized as follows
\begin{equation}\plabel{orthog}
\tilde{l}_m^{++} \tilde{l}^{m++} = 0,\quad
\tilde{l}_m^{--} \tilde{l}^{m--} = 0,\quad
\tilde{l}_m
\tilde{l}^{m\pm\pm} = 0,
\end{equation}
\begin{equation}\plabel{norm}
\tilde{l}_{m}^{--}
\tilde{l}^{m++} = 2,\quad
\tilde{l}_m
\tilde{l}^m = -1.
\end{equation}
Substituting these solutions into the Eqs. \p{MNconstr} and taking
account of the expressions \p{tMN} one finds
\begin{equation}\plabel{Mmm}
M \equiv M^{--}(\kpl, \eta^+, \oeta^+) =
(\overline{\tilde{D}}_+\oeta^+)
\tilde{\eta}^+\overline{\tilde{\eta}}^+
\tilde{l}_m^{--}(\tkpl)\tilde{\partial}_{++}\tilde{l}^m(\tkpl),
\end{equation}
\begin{equation}\label{Npp}
N \equiv M^{++}(\kml, \eta^-, \oeta^-) =
(\overline{\tilde{D}}_-\oeta^-)
\tilde{\eta}^-\overline{\tilde{\eta}}^-
\tilde{r}_m^{++}(\tkml)\tilde{\partial}_{--}\tilde{r}^m(\tkml)
\end{equation}
Now comparing these SFs with explicit form of the general solution of
the Eq. \p{nvn22sl}
\begin{eqnarray}\plabel{n22slsol}
e^{-2W(\kpml, \eta^{\pm})} &=& \frac{1}{2}r_m^{++}(\kml, \eta^-)
l^{--m}(\kpl, \eta^+), \\
\Psi^{--}_+(\kpl, \eta^+, \oeta^+) &=&
\frac{1}{2i}l_m^{--}(\kpl, \eta^+)D_+l^m(\kpl, \eta^+), \nn
\Psi^{++}_-(\kml, \eta^-, \oeta^-) &=&
\frac{1}{2i}r_m^{++}(\kml, \eta^-)D_-r^m(\kml, \eta^-), \nonumber
\end{eqnarray}
one can derives the following expressions for the corresponding SFs of
linear realization
\begin{equation}\plabel{2con}
M^{\pm\pm} = \Phi^{\pm}_{\pm}\Om^{\mp}\Psi^{\pm\pm}_{\mp},\qquad
\oD_{\pm}\Phi^{\pm}_{\pm} = 0, \qquad
D_{\pm}\Om^{\pm} = 1,
\end{equation}
\begin{equation}\label{OmPhisol}
\Phi^{\pm}_{\pm} = \overline{\tilde{D}}_{\pm}\oeta^{\pm},\qquad
\Om^{\pm} = (\tilde{D}_{\pm}\eta^{\pm})\tilde{\eta}^{\pm},
\end{equation}
\begin{eqnarray}\plabel{LRsol}
l_m^{\pm\pm,0}(\kpl, \eta^+) &=& \tilde{l}_m^{\pm\pm,0}(\tkpl), \\
r_m^{\pm\pm,0}(\kml, \eta^-) &=& \tilde{l}_m^{\pm\pm,0}(\tkml).
\nonumber
\end{eqnarray}

\section{Conclusion}
Thus, we have established that the $n=(2,2)$ generalization of complex
Liouville equation appropriated to the $N=2, D=4$ superstring is given by
the Eq. \p{nvn22sl} in which the auxiliary SFs $\Psi$ are subjected to
the constraints \p{sol} and \p{2con}. Then the general solution of this
equation can be given in terms of Lorentz harmonics \p{n22slsol}
which in one's turn are also restricted by the conditions \p{LRsol}.
Note, that in its own rights this fact actually means that the Eq.
\p{nvn22sl} proved to be exactly solvable as the corresponding bosonic string
equation does \cite{biku} but unlike to the bosonic case the corresponding
harmonic SFs becomes {\it essentially restricted} by the
constraints \p{LRsol} which provide the supersymmetric balance between
bosonic and fermionic degrees of freedom.
\footnote{It is obvious, that the same property of
$n=(1,1)$ super-Liouville equation \cite{bsv} can be derived from this one
with the
help of dimensional reduction from \p{n22slsol} and \p{LRsol}.}
It is worth mentioning that the first constraint in Eqs. \p{2con} implies
that the SFs $M$ are actually nilpotence $M^2 = 0$. From the theory
of spontaneously broken supersymmetries we know that such a type of
constraints leads directly to the nonlinear realizations of the underline
symmetries \cite{eiak}, \cite{rt}, in frame of which these constraints
could be solved explicitly in terms of the corresponding Goldstone
(super)fields. In the case under consideration we find the suitable
manifestly supercovariant solution \p{OmPhisol} in terms of the
Goldstone fermions of the nonlinear realization of $n=(2,2)$ superconformal
symmetry $\lambda^{\pm}(\kpml)$, $\ol^{\pm}(\kpml)$.

We are convinced that this
approach actually gives the universal way of deriving the equations of motion
as well as their solutions for the superstrings in the cases of higher
dimensions too, i.e. $D=6,10$.
In particular, the $N=2, D=6$ superstring is expected should be described
by the nonlinear realization of the $n=(4,4)$ supersymmetric WZNW
$\sigma$-model in which $W$
is replaced by the double-analytical SF $q^{(1,1)}$ representing twisted
multiplet in the harmonic $(4,4)$ superspace \cite{ikr}, \cite{ikl}.

We hope return to this question in a forthcoming publications.

\section*{Acknowledgments}

It is a great pleasure for me to express grateful to I.~Bandos, E.~Ivanov,
S.~Krivonos, A.~Pashnev and D.~Sorokin for interest to this work
and valuable discussions.

%       THIS DEFINES THE JOURNAL CITATIONS
\def\sep{~~}
\def\PRL #1 #2 #3 {{\em Phys. Rev. Lett.} {\bf#1} #2 (#3)}
\def\NPB #1 #2 #3 {{\em Nucl. Phys.} {\bf B #1} #2 (#3)}
\def\NPBFS #1 #2 #3 #4{{\em Nucl. Phys.} {\bf B #2} [FS#1] #3 (#4)}
\def\CMP #1 #2 #3 {{\em Commun. Math. Phys.} {\bf #1} #2 (#3)}
\def\PRD #1 #2 #3 {{\em Phys. Rev.} {\bf D #1} #2 (#3)}
\def\PRev #1 #2 #3 {{\em Phys. Rev.} {\bf #1} #2 (#3)}
\def\PLA #1 #2 #3 {{\em Phys. Lett.} {\bf #1A} #2 (#3)}
\def\PLB #1 #2 #3 {{\em Phys. Lett.} {\bf B #1} #2 (#3)}
\def\JMP #1 #2 #3 {{\em J. Math. Phys.} {\bf #1} #2 (#3)}
\def\PTP #1 #2 #3 {{\em Prog. Theor. Phys.} {\bf #1} #2 (#3)}
\def\SPTP #1 #2 #3 {{\em Suppl. Prog. Theor. Phys.} {\bf #1} #2 (#3)}
\def\AoP #1 #2 #3 {{\em Ann. of Phys.} {\bf #1} #2 (#3)}
\def\PNAS #1 #2 #3 {{\em Proc. Natl. Acad. Sci. USA} {\bf #1} #2 (#3)}
\def\RMP #1 #2 #3 {{\em Rev. Mod. Phys.} {\bf #1} #2 (#3)}
\def\PR #1 #2 #3 {{\em Phys. Reports} {\bf #1} #2 (#3)}
\def\AoM #1 #2 #3 {{\em Ann. of Math.} {\bf #1} #2 (#3)}
\def\UMN #1 #2 #3 {{\em Usp. Mat. Nauk} {\bf #1} #2 (#3)}
\def\FAP #1 #2 #3 {{\em Funkt. Anal. Prilozheniya} {\bf #1} #2 (#3)}
\def\FAaIA #1 #2 #3 {{\em Functional Analysis and Its Application}
  {\bf #1} #2 (#3)}
\def\BAMS #1 #2 #3 {{\em Bull. Am. Math. Soc.}
  {\bf #1} #2 (#3)}
\def\TAMS #1 #2 #3 {{\em Trans. Am. Math. Soc.} {\bf #1} #2 (#3)}
\def\InvM #1 #2 #3 {{\em Invent. Math.} {\bf #1} #2 (#3)}
\def\LMP #1 #2 #3 {{\em Letters in Math. Phys.} {\bf #1} #2 (#3)}
\def\IJMPA #1 #2 #3 {{\em Int. J. Mod. Phys.} {\bf A #1} #2 (#3)}
\def\AdM #1 #2 #3 {{\em Advances in Math.} {\bf #1} #2 (#3)}
\def\RMaP #1 #2 #3 {{\em Reports on Math. Phys.} {\bf #1} #2 (#3)}
\def\IJM #1 #2 #3 {{\em Ill. J. Math.} {\bf #1} #2 (#3)}
\def\APP #1 #2 #3 {{\em Acta Phys. Polon.} {\bf #1} #2 (#3)}
\def\TMP #1 #2 #3 {{\em Theor. Mat. Phys.} {\bf #1} #2 (#3)}
\def\JPA #1 #2 #3 {{\em J. Physics} {\bf A#1} #2 (#3)}
\def\JPG #1 #2 #3 {{\em J. Physics} {\bf G#1} #2 (#3)}
\def\JSM #1 #2 #3 {{\em J. Soviet Math.} {\bf #1} #2 (#3)}
\def\MPLA #1 #2 #3 {{\em Mod. Phys. Lett.} {\bf A #1} #2 (#3)}
\def\JETP #1 #2 #3 {{\em Sov. Phys. JETP} {\bf #1} #2 (#3)}
\def\JETPL #1 #2 #3 {{\em Sov. Phys. JETP Lett.} {\bf #1} #2 (#3)}
\def\PHSA #1 #2 #3 {{\em Physica} {\bf A #1} #2 (#3)}
\def\CQG #1 #2 #3 {{\em Class. Quantum Grav.} {\bf #1} #2 (#3)}
\def\SJNP #1 #2 #3 {{\em Sov. J. Nucl. Phys. (Yadern.Fiz.)} {\bf #1} #2 (#3)}
\def\JEPAN #1 #2 #3 {{\em J. Elem. Part. Atom. Nucl.} {\bf #1} #2 (#3)}

\end{document}